\documentclass[10pt,twocolumn]{article}
\usepackage{isqcmc2025} 

\usepackage{graphicx,url}

\usepackage[utf8]{inputenc}
\usepackage{amsmath,physics}
\usepackage{tikz,yquant,amsfonts}
\useyquantlanguage{groups}
\usetikzlibrary{quotes, calc}

\usepackage{anonymize} 
\usepackage{listings}

\lstdefinelanguage{OPENQASM}{
morekeywords={q, c}, 
morekeywords=[2]{OPENQASM, include}, 
emph={h,cx,qreg,creg,->}, 
morekeywords=[3]{measure}, 
morekeywords=[4]{barrier}, 
sensitive=true,
morecomment=[l]{//}, 
morestring=[b]",
literate={->}{{\textbf{\color{codeemph2}{$\to$}}}}1 
}

\usepackage{orcidlink}

\title{Processing through encoding: Quantum circuit approaches for point-wise multiplication and convolution}

\author{\anonymize{Andreas Papageorgiou, }\inst{1}\thanks{This project has received funding from \anonymize{the European Union’s} \anonymize{HORIZON research and innovation programme} \anonymize{HORIZON-WIDERA-2022-TALENTS-01 under grant agreement No. 101087126}.}\(\,\,^{\dagger}\) \orcidlink{0009-0001-9164-0980} \and
        \anonymize{Paulo Vitor Itabora\'i}\inst{1}\inst{2}\(^{*}\) \orcidlink{0000-0002-4956-2958} \and \anonymize{Kostas Blekos}\inst{1}\(^{*}\) \orcidlink{0000-0002-6777-2107} \and \anonymize{Karl Jansen}\inst{1}\inst{2}\(^{*}\) \orcidlink{0000-0002-1574-7591}}

\address{\anonymize{Computation-Based Science and Technology Research Center -- The Cyprus Institute}\\
         \anonymize{20 Kavafi Street - 2121 - Nicosia - Cyprus}
         \nextinstitute
         \anonymize{Center for Quantum Technology and Applications -- Deutsches Elektronen-Synchrotron DESY} \\
         \anonymize{Platanenallee 6 - 15738 - Zeuthen - Brandenburg - Germany}
         \email{\(\dagger\) \anonymize{an.papageorgiou@cyi.ac.cy}}
}

\begin{document}

\maketitle

\begin{abstract}
This paper introduces quantum circuit methodologies for pointwise multiplication and convolution of complex functions, conceptualized as ``processing through encoding''.
Leveraging known techniques,
we describe an approach where multiple complex functions are encoded onto auxiliary qubits.
Applying the proposed scheme for two functions $f$ and $g$, their pointwise product $f(x)g(x)$ is shown to naturally form as the coefficients of part of the resulting quantum state.
Adhering to the convolution theorem, we then demonstrate how the convolution $f*g$ can be constructed.
Similarly to related work, this involves the encoding of the Fourier coefficients $\mathcal{F}[f]$ and $\mathcal{F}[g]$, which facilitates their pointwise multiplication, followed by the inverse Quantum Fourier Transform.
We discuss the simulation of these techniques, their integration into an extended \verb|quantumaudio| package for audio signal processing, and present initial experimental validations.
This work offers a promising avenue for quantum signal processing, with potential applications in areas such as quantum-enhanced audio manipulation and synthesis.
\end{abstract}

\section{Introduction}\label{sec:introduction}
Since its conception, the use of quantum computation has found potential uses across numerous fields of research.
With its growth, pioneering methods for its artistic use have also been explored, 
especially for music and visual arts \cite{mirandaQuantumComputingArts2022,mirandaQuantumComputerMusic2022,miranda2025qcmbook}.
This further includes the codification and audification of quantum phenomena for the production of musical compositions \cite{itaborai2023towards}.
To this purpose, various methodologies which encode arbitrary one-dimensional signals on the wavefunction of qubits have been studied and developed. Some of these methods are compiled into a comprehensive survey carried by Itabora\'i and Miranda \cite{Itaborai2022,itaborai2023towards}.
In this work, we seek to examine how future quantum tools and methodologies might be used to generate, synthesise, and process arbitrary waveforms by leveraging the structure of signal preparation circuits.

In Section~\ref{sec:encoding_methodology}, we detail the encoding of arbitrary complex functions via a probabilistic-based approach with superposition indexing.
In Section~\ref{sec:algo_pointwise_multiplication}, we demonstrate how encoding two complex signals on a wavefunction allows for the computation of their pointwise multiplication as part of the resulting state.
Building upon this, Section~\ref{sec:algo_convolution} discusses how this method can be extended to encode the convolution of arbitrary functions, by leveraging the pointwise multiplication of their Fourier coefficients, followed by an inverse Quantum Fourier Transform.
The result comes in accordance with similar techniques such as the ones presented by 
Motlagh and Wiebe \cite{PRXQuantum.5.020368} as well as  Nair et al. \cite{Nair2025}.
In Section~\ref{sec:simulation_audio}, we describe the integration of this approach into the \verb|quantumaudio| \cite{moth_quantum_and_collaborators_2024_14025598} package for audio signal processing, present simulation results for pointwise multiplication, and comment on practical considerations.
Finally, Section~\ref{sec:discussion} provides a concluding discussion of the presented methods and outlines future work.

\section{Encoding Complex Functions on Quantum States}\label{sec:encoding_methodology}
To construct a quantum state that represents a complex function we combine two similar audio encodings: 
Single Qubit Probability Amplitude Modulation (SQPAM) and Single Qubit Probability Phase Modulation 
(SQPPM)\cite{itaborai2023towards}. These can be seen as 1D equivalents of 
image encodings described in \cite{le2011flexible} and subsequently its sample-addressable audio counterpart \cite{Itaborai2022}.

\subsection{General Encoding Principle}\label{ssec:general_encoding}
For $N=2^n$, we will be using an integer indexed basis
\begin{align}
    \ket{0}   = \begin{pmatrix} 1 \\ 0 \\ \vdots \\ 0 \end{pmatrix}, &&
    \ket{1}   = \begin{pmatrix} 0 \\ 1 \\ \vdots \\ 0 \end{pmatrix}, && 
    \cdots &&
    \ket{N-1} = \begin{pmatrix} 0 \\ 0 \\ \vdots \\ 1 \end{pmatrix}.
\end{align}
An arbitrary complex function $f : \{0,1,\dots,N-1\} \mapsto \mathbb{C}$
can then be encoded as a wavefunction $\ket{f}$ of an
$n$-qubit register, such that up to normalization:
\begin{equation}
    \ket{f} = \sum_{x=0}^{N-1}f(x)\ket{x}.
    \label{eq:target-encoding}
\end{equation}
To achieve this, we can first create a uniform superposition over all inputs
by applying a Hadamard gate on every qubit
of a quantum register $q$, of size $n$:
\begin{equation}
    \ket{\psi_0} =
    H^{\otimes n} \ket{0}^{\otimes n} =
    \frac{1}{\sqrt{N}} \sum_{x = 0}^{N-1} \ket{x}
\end{equation}
One way to encode $f$ would then be to use an additional
qubit, labeled $t$, and construct transformations of the form
\begin{equation}
    \begin{tikzpicture}[baseline=-0.925\baselineskip]
        \begin{yquant}
            qubit {} q;
            box {$\rho_f(x)$} q;
        \end{yquant}
    \end{tikzpicture}
    =
    \begin{bmatrix}
        f(x)              & \cdot \\
        \cdot & \cdot
    \end{bmatrix},
\end{equation}
where entries marked with $\cdot$ must guarantee $\rho$
is unitary.
\begin{figure}
    \centering
    \begin{tikzpicture}
        \setlength\arraycolsep{1pt}
        \begin{yquant}
            qubit {$q_0: \ket{0}$} x;
            qubit {$q_1: \ket{0}$} x[+1];
            qubit {$\vdots$} d;
            qubit {$q_{n-1}: \ket{0}$} x[+1];
            qubit {$t: \ket{0}$} t;
            discard d;
            box {$H$} x;
            box {$\rho_f(0)$} t | ~x;
            box {$\rho_f(1)$} t | x[0], ~x[1], x[2]; 
            text {$\cdots$} -;
            box {$\rho_f(N-1)$} t | x;
        \end{yquant} 
    \end{tikzpicture}
    \caption{
        Schematic for encoding a complex function $f$.
        All qubits are initially in state $\ket{0}$.
        An $n$-qubit register is placed in uniform superposition
        over all its computational basis states.
        Value setting operations $\rho_f(x)$ are applied to the
        ancilla $t$.
    }
    \label{fig:circuit-multicontrol}
\end{figure}
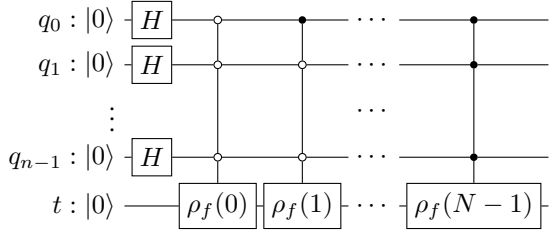
For each input $x$, its corresponding evaluation $f(x)$
is a complex number $re^{i\theta}$, whose
magnitude $|f(x)| = r$ and phase $\arg(f(x)) = \theta$.
Figure~\ref{fig:circuit-multicontrol} summarizes
the encoding of $f(x)$
by applying a corresponding $\rho_f(x)$ transformation on qubit $t$,
conditioned on $q$ being in each basis state $\ket{x}$.

The part of the resulting state where $t$ is $\ket{0}$ will
contain the encoded function, since
\begin{equation}
    \begin{gathered}
        \rho_f(x)\ket{0} = f(x)\ket{0} + \widetilde{f(x)}\ket{1},\\
        \text{ for some $\widetilde{f}$ satisfying } |f(x)|^2 +
        \left|\widetilde{f(x)}\right|^2 = 1.
    \end{gathered}
\end{equation}
Transformations $\rho_f(x)$ can be constructed
by separately encoding the magnitude and phase information
of $f(x)$ with gates labeled $\mu_f(x)$ and $\phi_f(x)$ respectively,
such that:
\begin{equation}
    \label{eq:rho}
    \begin{tikzpicture}[baseline=-0.925\baselineskip]
        \begin{yquant}
            qubit {} q_anc; 
            box {$\rho_f(x)$} q_anc;
        \end{yquant}
    \end{tikzpicture}
    =
    \begin{tikzpicture}[baseline=-0.925\baselineskip]
        \begin{yquant}
            qubit {} q_anc; 
            box {$\mu_{f}(x)$} q_anc;
            box {$\phi_{f}(x)$} q_anc;
        \end{yquant}
    \end{tikzpicture}
    .
\end{equation}
These are described in more detail in the sections
below.

\subsection{Magnitude Encoding}\label{ssec:magnitude_encoding}
After renormalizing $f$ such that $|f(x)| \in [0,1)$,
for each input $x$ we can construct a rotation:
\begin{equation}\label{eq:magnitude_encoding}
    \begin{gathered}
        \begin{tikzpicture}[baseline=-0.75\baselineskip]
            \begin{yquant}
                qubit {} q_anc; 
                box {$\mu_f(x)$} q_anc;
            \end{yquant}
        \end{tikzpicture}
        = R_Y(2\theta_x) = \begin{bmatrix}
            \cos(\theta_x) &           -\sin(\theta_x) \\
            \sin(\theta_x) & \phantom{-}\cos(\theta_x)
        \end{bmatrix}, \\ \text{where} \hspace{1cm}
        \theta_x = \arccos(|f(x)|).
    \end{gathered}
\end{equation}
The resulting range of $\arccos$ is taken as $\left[0,\frac{\pi}{2}\right)$.
Thus $\cos(\theta_x) = |f(x)|$.
Denoting $\widetilde{f(x)} = \sqrt{1-|f(x)|^2}$,
the state after applying these controlled magnitude encodings, as shown in the 
first part of Figure~\ref{fig:circuit-encode}, will be:
\begin{equation}
    \ket{\psi_1} = \frac{1}{\sqrt{N}}\sum_{x =0}^{N-1} \ket{x}\otimes\left(
        \left|f(x)\right|\ket{0} +
        \widetilde{f(x)}\ket{1}
    \right).
\end{equation}
The magnitude $|f(x)|$ is therefore encoded in the part of the state where qubit $t$ is $\ket{0}$.

\subsection{Phase Encoding}\label{ssec:phase_encoding}
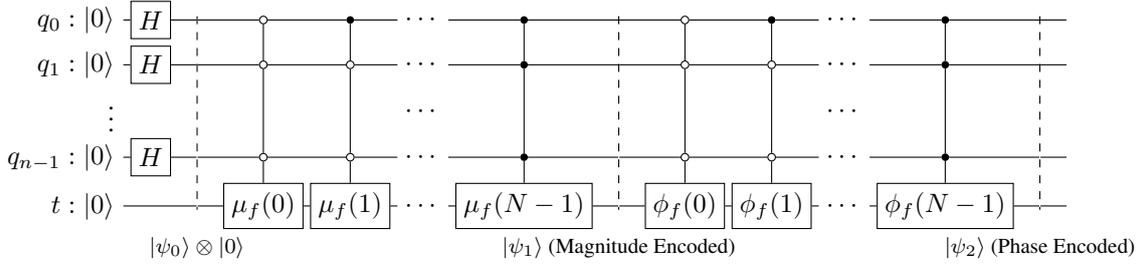
\begin{figure*}[htbp] 
    \centering
    \begin{tikzpicture}
        \begin{yquant}
            qubit {$q_0: \ket{0}$} x_reg; 
            qubit {$q_1: \ket{0}$} x_reg[+1];
            qubit {$\vdots$} d;
            qubit {$q_{n-1}: \ket{0}$} x_reg[+1];
            qubit {$t: \ket{0}$} t_anc; 
            discard d;
            box {$H$} x_reg;
            [name=b0] barrier (-);
            box {$\mu_f(0)$} t_anc | ~x_reg;
            box {$\mu_f(1)$} t_anc | x_reg[0], ~x_reg[1], x_reg[2]; 
            text {$\cdots$} -;
            box {$\mu_f(N-1)$} t_anc | x_reg;
            [name=b1] barrier (-);
            box {$\phi_f(0)$} t_anc | ~x_reg;
            box {$\phi_f(1)$} t_anc | x_reg[0], ~x_reg[1], x_reg[2]; 
            text {$\cdots$} -;
            box {$\phi_f(N-1)$} t_anc | x_reg;
            [name=b2] barrier (-);
        \end{yquant}
        \node at ($(b0.south) + (0, -0.5)$) {\footnotesize $\ket{\psi_0}\otimes \ket{0}$};
        \node at ($(b1.south) + (0, -0.5)$) {\footnotesize $\ket{\psi_1}$ (Magnitude Encoded)};
        \node at ($(b2.south) + (0, -0.5)$) {\footnotesize $\ket{\psi_2}$ (Phase Encoded)};
    \end{tikzpicture}
    \caption{Multi-step encoding of a complex function $f$ using $n+1$ qubits.
    A uniform superposition over all inputs, $\ket{\psi_0}$
    is tensored with an ancilla qubit in ground state $\ket{0}$.
    Controlled $\mu_f(x)$ and $\rho_f(x)$ gates then encode the magnitude and
    phase information of $f(x)$ into states $\ket{\psi_1}$ and $\ket{\psi_2}$ 
    respectively.}
    \label{fig:circuit-encode}
\end{figure*}

To encode the phase $\arg(f(x))$ of the complex function $f(x)$, for each
input $x$ we construct phase transformations:
\begin{equation}
    \begin{gathered}
        \begin{tikzpicture}[baseline=-0.7\baselineskip]
            \begin{yquant}
                qubit {} q_anc; 
                box {$\phi_f(x)$} q_anc;
            \end{yquant}
        \end{tikzpicture}
        = \begin{pmatrix}
            e^{i\arg(f(x))} & 0\\
            0 & 1 
        \end{pmatrix}.
    \end{gathered}
\end{equation}
These are applied as a series of multi-controlled phase gates,
conditioned on the register $q$ being in each state $\ket{x}$,
as show in the second part of Figure~\ref{fig:circuit-encode}.

\subsection{Combined Amplitude and Phase Encoding}\label{sec:combined_encoding}
The phase gate $\phi_f(x)$ acts as $P(\arg(f(x)))$ on $\ket{0}$ and identity on $\ket{1}$. The term $\widetilde{f(x)}$ from magnitude encoding therefore picks up
no additional phase. So, $f(x)$ is correctly encoded in the
amplitude of the $\ket{x}\ket{0}$ component:
\begin{equation}
    \ket{\psi_2} = \frac{1}{\sqrt{N}}\sum_{x =0}^{N-1} \ket{x}\otimes\left(
        f(x)\ket{0} + \widetilde{f(x)}\ket{1}
    \right)
\end{equation}
We further note that due to the commutativity of the
multi-controlled gates, the two encodings can be interleaved such
that controlled operations $\mu_f(x)$ and $\phi_f(x)$ are performed
successively, conditioned on each respective basis state $\ket{x}$.
Using Equation~\ref{eq:rho}, 
successive gates $\mu_f(x)$ and $\phi_f(x)$ can then be joined to form
a unified value setting operation, as shown in 
Figure~\ref{fig:circuit-multicontrol}.

\section{Quantum Algorithm for Pointwise Multiplication}\label{sec:algo_pointwise_multiplication}
\begin{figure*}[tbp] 
    \centering
    \setlength\arraycolsep{1pt}
    \begin{tikzpicture}
        \begin{yquant}
            qubit {$q_0: \ket{0}$} x_reg; 
            qubit {$q_1: \ket{0}$} x_reg[+1];
            qubit {$\vdots$} d;
            qubit {$q_{n-1}: \ket{0}$} x_reg[+1];
            qubit {$t_f: \ket{0}$} tf_anc; 
            qubit {$t_g: \ket{0}$} tg_anc; 
            discard d;
            box {$H$} x_reg;
           [name=b0] barrier (-);
            box {$\rho_f(0)$} tf_anc | ~x_reg;
            box {$\rho_f(1)$} tf_anc | x_reg[0], ~x_reg[1], x_reg[2]; 
            text {$\cdots$} -;
            box {$\rho_f(N-1)$} tf_anc | x_reg;
            [name=b1] barrier (-);
            box {$\rho_g(0)$} tg_anc | ~x_reg;
            box {$\rho_g(1)$} tg_anc | x_reg[0], ~x_reg[1], x_reg[2]; 
            text {$\cdots$} -;
            box {$\rho_g(N-1)$} tg_anc | x_reg;
            [name=b2] output {\footnotesize{$\begin{matrix}
                    \frac{1}{\sqrt{N}}\sum_x f(x)g(x)\ket{x}\ket{00}                         & + \\
                    \frac{1}{\sqrt{N}}\sum_x f(x)\widetilde{g(x)}\ket{x}\ket{01}             & + \\
                    \frac{1}{\sqrt{N}}\sum_x \widetilde{f(x)}g(x)\ket{x}\ket{10}             & + \\
                    \frac{1}{\sqrt{N}}\sum_x \widetilde{f(x)}\widetilde{g(x)}\ket{x}\ket{11} 
                \end{matrix}
            $}} (-);
        \end{yquant}
        \node at ($(b0.south) + (0, -0.5)$) {\footnotesize $\ket{\psi_0}\otimes \ket{00}$};
        \node[below=1pt] at ($(b0.south west)!0.5!(b1.south east) + (0,-0.25)$) {Encode $f$ on $t_f$};
        \node[below=1pt] at ($(b1.south west)!0.5!(b2.south east) + (-1.75,-0.25)$) {Encode $g$ on $t_g$};
        \node at ($(b2.north) + (0, -0.5)$) {\footnotesize $\ket{\psi_3}=$};
    \end{tikzpicture}
    \caption{Pointwise multiplication of complex functions $f$ and $g$ using $n+2$ qubits. After preparing register $q$ in superposition, $f$ is encoded using ancilla $t_f$, and $g$ is encoded using ancilla $t_g$. The desired product $f(x)g(x)$ is associated with the $\ket{00}$ state of the ancillae $t_f t_g$.}
    \label{fig:circuit-multiply}
\end{figure*}
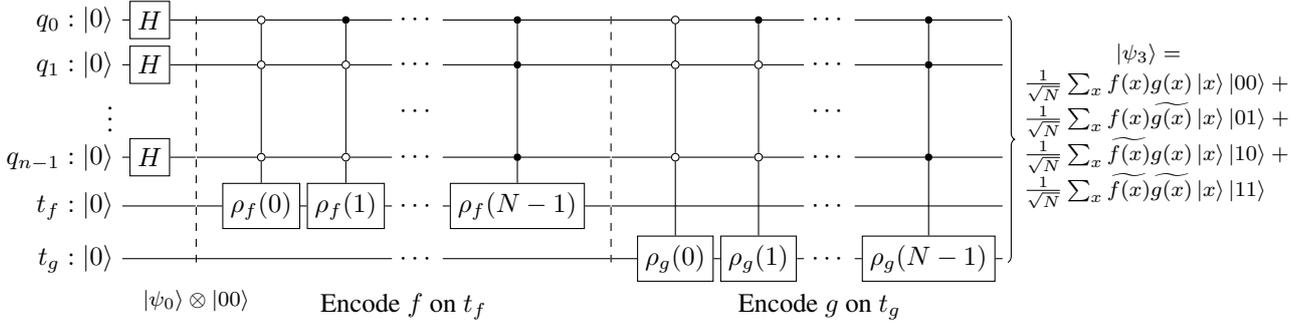

Given a second complex function $g : \{0,\dots,N-1\} \mapsto \mathbb{C}$,
renormalized such that $|g(x)| \in [0,1)$,
we aim to obtain a quantum state proportional to
its pointwise product with $f$:
\begin{equation}
    \ket{f \cdot g} = \sum_{x=0}^{N-1} f(x)g(x)\ket{x}.
    \label{eq:multiply_target} 
\end{equation}
To achieve this, we use the $n$-qubit register $q$ (initially in $\ket{\psi_0}$) and two ancilla qubits, $t_f$ and $t_g$, initialized to $\ket{00}$.
We apply the encoding procedure described in Section~\ref{sec:encoding_methodology} for function $f$ using $t_f$ as its ancilla, and repeat for function $g$ using $t_g$ as its ancilla.
The circuit is shown in Figure~\ref{fig:circuit-multiply}. After these operations, the resulting quantum state will be:
\begin{equation}\label{eq:qpte_expanded}
    \begin{split}
        \ket{\psi_3} = \frac{1}{\sqrt{N}}\sum_{x=0}^{N-1}\ket{x} \otimes \Big(
                 & f(x)g(x) \ket{00}_{t_f t_g} \\
            {}+{}& f(x)\widetilde{g(x)} \ket{01}_{t_f t_g} \\
            {}+{}& \widetilde{f(x)}g(x) \ket{10}_{t_f t_g} \\
            {}+{}& \widetilde{f(x)}\widetilde{g(x)} \ket{11}_{t_f t_g}
        \Big) .
    \end{split}
\end{equation}
If upon measurement, the ancilla register $t_f t_g$ collapses to $\ket{00}$, the post-selected state of qubit register $q$ will be (up to normalization): $\sum_x f(x)g(x)\ket{x}$, which is the desired pointwise product $\ket{f \cdot g}$. The probability of measuring $\ket{00}$ is $\frac{1}{N}\sum_x |f(x)g(x)|^2$.

\section{Quantum Algorithm for Convolution}\label{sec:algo_convolution}
Thereafter, the convolution of two discrete functions $f, g: \{0, \dots, N-1\} \mapsto \mathbb{C}$ is typically defined for periodic functions or requires zero-padding. Assuming appropriate handling (e.g., functions are $N$-periodic or zero-padded to length $M \ge 2N-1$ to avoid wraparound effects for linear convolution), the discrete convolution is:
\begin{equation}
    (f*g)(k) = \sum_{j=0}^{M-1} f(j)g(k - j \pmod M).
\end{equation}
The result $(f*g)$ is a function over $\{0, \dots, M-1\}$. If $M=2^m$, it can be encoded in an $m$-qubit register:
\begin{equation}\label{eq:convolved_target}
    \ket{f * g} = \sum_{k=0}^{M-1} (f * g)(k)\ket{k}.
\end{equation}
We can construct this state by combining the pointwise multiplication technique (Section~\ref{sec:algo_pointwise_multiplication}) with the Quantum Fourier Transform (QFT)\cite{nielsen2002quantum}.

\subsection{Convolution via the Convolution Theorem}\label{ssec:convolution_theorem}
Using the $M^{\text{th}}$ root of unity $\omega = e^{-2\pi i/M}$, 
the  Discrete Fourier Transform (DFT) of a function $f$ (of length $M$)
is:
\begin{equation}
    \mathcal{F}[f](x) = \widehat{f}(x) = \sum_{y=0}^{M-1} f(y) \omega^{xy}.
\end{equation}
The QFT is the DFT's quantum analogue, transforming:
\begin{equation}
    \ket{\widehat{f}} = \text{QFT}\ket{f} = \sum_{x=0}^{M-1} \widehat{f}(x)\ket{x}.
\end{equation}
The Convolution Theorem states that the DFT of a convolution is the pointwise product of the DFTs:
\begin{align}
    \mathcal{F}[f*g]_l = \mathcal{F}[f]_l \cdot \mathcal{F}[g]_l
    && \text{or} &&
    \widehat{(f*g)}_l = \widehat{f}_l \cdot \widehat{g}_l.
\end{align}
Therefore, $f*g = \mathcal{F}^{-1}[\widehat{f} \cdot \widehat{g}]$.
The state of Equation~\ref{eq:convolved_target} can thus be reached following these steps:
\begin{enumerate}
    \item Evaluate $\widehat{f}$ and $\widehat{g}$ classicaly.
    This typically means $f$ and $g$ are first zero-padded to length $M$ (e.g., 
    $M=2N$ if $N$ is a power of 2, requiring $n+1$ qubits for the index register).
    \item Use the method from Section~\ref{sec:algo_pointwise_multiplication} to 
    encode $\widehat{f}$ and $\widehat{g}$ using ancilla qubits $t_{f}$ and $t_{g}$
    respectively. Post-selecting on the part of the state where ancillas are $\ket{00}$,
    the state of the $m$-qubit register $q$ will be, up to normalization, the pointwise
    product of the Fourier coefficients: $\ket{\widehat{f}\cdot\widehat{g}}$.
    \item Apply the inverse Quantum Fourier Transform (QFT$^\dagger$) to the index register $q$,
    The resulting state, up-to normalization matches Equation~\ref{eq:convolved_target}
\end{enumerate}
The QFT and its inverse QFT$^\dagger$ are to known to have efficient implementations on a quantum computer, using $O(m^2)$ gates for $m$ qubits \cite{nielsen2002quantum}.

\subsection{Optimizing Classical Computation Steps}\label{ssec:optimizing_classical_convolution}
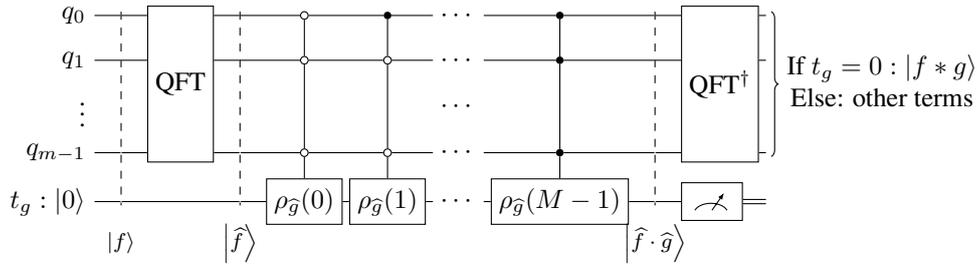
\begin{figure*}[htbp] 
    \centering
    \begin{tikzpicture}
        \begin{yquant} 
            qubit {$q_0$} x_q; 
            qubit {$q_1$} x_q[+1];
            qubit {$\vdots$} d_q;
            qubit {$q_{m-1}$} x_q[+1];
            qubit {$t_g:\ket{0}$} t_anc_g; 
            discard d_q;
            [name=b0] barrier (-);
            box {QFT} (x_q, d_q); 
            [name=b1] barrier (-);
            box {$\rho_{\widehat{g}}(0)$} t_anc_g | ~x_q;
            box {$\rho_{\widehat{g}}(1)$} t_anc_g | x_q[0], ~x_q[1], x_q[2]; 
            text {$\cdots$} -;
            box {$\rho_{\widehat{g}}(M-1)$} t_anc_g | x_q;
            [name=b2] barrier (-);
            box {QFT$^\dagger$} (x_q, d_q); 
            measure t_anc_g;
            output {$
                \begin{matrix}
                   \text{If } t_g=0: \ket{f * g} \\
                   \text{Else: other terms}
                \end{matrix}
            $} (x_q,d_q);
        \end{yquant}
        \node at ($(b0.south) + (0, -0.5)$) {\footnotesize $\ket{f}$};
        \node at ($(b1.south west)!0.5!(b1.south east) + (0,-0.5)$) {\footnotesize $\ket{\widehat{f}}$};
        \node at ($(b2.south west)!0.5!(b2.south east) + (0,-0.5)$) {\footnotesize
            $\ket{\widehat{f}\cdot\widehat{g}}$
        };
    \end{tikzpicture}
    \caption{Convolution of an initially encoded function $f$ with a kernel $g$ (whose Fourier Transform $\widehat{g}$ is encoded). State $\ket{f}$ is transformed to Fourier domain $\ket{\widehat{f}}$. Then $\widehat{g}$ is encoded using ancilla $t_g$. After post-selecting $t_g=\ket{0}$, an inverse QFT yields $\ket{f*g}$.}
    \label{fig:circuit-convolution}
\end{figure*}

The above procedure requires classical computation of $\widehat{f}$ and $\widehat{g}$. We can reduce some classical computation if one function, say $f$, is already encoded in the time/spatial domain as
Equation \ref{eq:target-encoding}.
The procedure, illustrated in Figure~\ref{fig:circuit-convolution}, then becomes:
\begin{enumerate}
    \item Start with state $\ket{f}\ket{0}_{t_f}\ket{0}_{t_g}$ (or prepare $\ket{f}$ on $q$ and initialize ancillae). If $f$ is given classically, encode it first. Assume $f$ is already appropriately zero-padded.
    \item Apply QFT to the register $q$ holding $\ket{f}$. This transforms it to $\ket{\widehat{f}}$. The state is now $\ket{\widehat{f}}\ket{0}_{t_f}\ket{0}_{t_g}$.
    \item Classically compute $\widehat{g}$ (the Fourier transform of the convolution kernel $g$).
    \item Using controlled $\rho_{\widehat{g}}(x)$ gates, encode $\widehat{g}$ onto ancilla $t_g$, conditioned on each basis of register $q$.
    The operations are controlled $\rho_{\widehat{g}}(l)$ gates applied to $t_g$.
    The state becomes $\sum_l \widehat{f}_l \ket{l} \otimes (\widehat{g}_l\ket{0} + \widetilde{\widehat{g}_l}\ket{1})_{t_g}$.
    \item Post-select on $t_g = \ket{0}$. The state of $q$ becomes $\sum_l \widehat{f}_l \widehat{g}_l \ket{l}$.
    \item Apply QFT$^\dagger$ to register $q$. This yields $\ket{f*g}$.
\end{enumerate}
This approach still requires classical computation of $\widehat{g}$. If $g$ (and thus $\widehat{g}$) has a structure allowing for efficient quantum state preparation (e.g., $g$ is a simple FIR filter), then the classical burden can be further reduced. This scheme aligns with related work \cite{PRXQuantum.5.020368, Nair2025} where operations are often analyzed in the Fourier basis.
It further provides a concrete construction of quantum circuits
that encode convolutions of two arbitrary functions using only 2 additional qubits.

\section{Simulation and Application to Audio Signals}\label{sec:simulation_audio}

In pursuit of musical applications of the methodology so far described,
we begin by employing an audification of the pointwise multiplication circuit of Section \ref{sec:algo_pointwise_multiplication}. To operate with audio signals, we modified and extended the open-source \verb|quantumaudio| package \cite{moth_quantum_and_collaborators_2024_14025598} to incorporate the as here referred "Quantum Processing Through Encoding" (QPTE) approach.

Our implementation extends the SQPAM encoding scheme, to include the magnitude and phase encoding (Section~\ref{sec:encoding_methodology}), along with utility functions for building the pointwise multiplication circuit (Section~\ref{sec:algo_pointwise_multiplication}),
and metadata handling within the package's structure.

\subsection{Methodology}\label{sssec:exp_methodology}
In coefficient-based quantum encodings, audio signals are typically mapped from $[-1,1)$ to $[0,1)$.
However, such shifting would alter the outcome of a true pointwise multiplication. For our experiments, input signals were assumed to be strictly in the positive domain (e.g., scaled to $[0,1)$).

The pointwise multiplication of two such positive signals, $f(x)$ and $g(x)$, is then computed by encoding 
them onto two ancillae qubts ($t_f, t_g$) as per Section~\ref{sec:algo_pointwise_multiplication}. 
All four components of Equation~\ref{eq:qpte_expanded} are extracted from each simulation run,
effectively producing a quadraphonic output. 

These cross-terms retain inter-relational information from the original inputs and hence outline an initial
interest for studying its musical potential.
The audification of all resulting signals provides further qualitative validation of the process.

To keep computational resources to a minimum, the quantum circuits simulated were comprised of 5 qubits.
Audio signals were split and encoded in chunks of $2^3 = 8$ samples. Measures of state fidelity and
root mean square deviation were then taken to compare the accuracy of the sampled wavefunction
against the ideal pointwise product of the encoded inputs. The
processing of each audio chunk was then distributed into a
parallelised task within a cluster environment providing linear scaling depending on the number of parallel computational units. This allowed us to compute substantially larger audio files, compared to previous literature \cite{itaborai2023towards}, maintaining
an acceptable accuracy and simulation time (Section \ref{sssec:exp_results}).

\subsection{Results and Analysis}\label{sssec:exp_results}
\begin{figure}[h]
\centering
  \includegraphics[width=.45\textwidth]{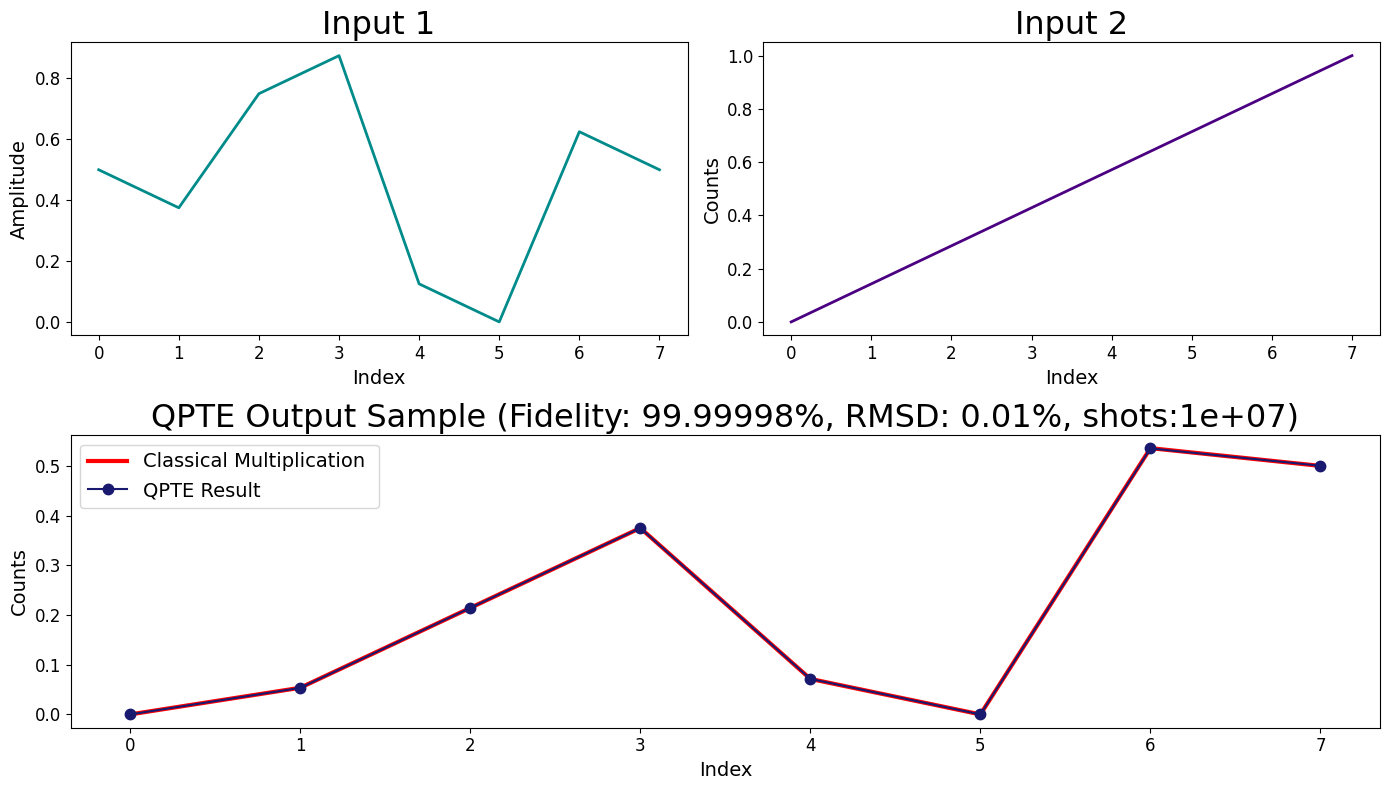}
\caption{Sample experiment for computing the Point-wise multiplication of two signals using the \texttt{quantumaudio} package}
\label{fig:pointwise_results1}
\end{figure}
\noindent
Initial simulations demonstrate a successful pointwise multiplication, with deviations attributable to finite sampling (shot noise) in the quantum simulation. Table \ref{tab:rmsd_fidelity} displays a shot scaling test for a single experiment.
It is observed that the fidelity approaches 100\%, faster than the RMSD reaches 0\%. For this case, a near perfect reconstruction (i.e, with negligible audible white noise) was achieved for RMSD $<$ 0.01\%.

\begin{table}[h]
\centering
\scriptsize
\setlength{\tabcolsep}{3pt}
\begin{tabular}{@{}lccccccc@{}}
\hline
 \textbf{Shots} & \textbf{1e1} & \textbf{1e2} & \textbf{1e3} & \textbf{1e4} & \textbf{1e5} & \textbf{1e6} & \textbf{1e7} \\
\hline
\textbf{RMSD (\%)}     & 20.59 & 6.16 & 1.39 & 0.39 & 0.10 & 0.04 & 0.01 \\
\textbf{Fidelity (\%)} & 52.639 & 96.087 & 99.617 & 99.978 & 99.998 & 99.99985 & 99.99998 \\
\hline
\end{tabular}
\caption{QPTE Shot scaling test for an experiment with 8 samples (Figure \ref{fig:pointwise_results1}).}\label{tab:rmsd_fidelity}
\end{table}

Preliminary experiments comparing a serial run (on a laptop $\sim$856 seconds) and a paralellised run with 32 cores (on the cluster $\sim$37 seconds) verify a speedup of our simulation times proportional to the computational resources.
Source code can be found in \cite{papageorgiou_2025_17158544}.

\section{Discussion}\label{sec:discussion}
To process information such as image or sound, quantum methods often require the encoding of arbitrary functions as wavefunctions of qubits.
Focusing on 1-dimensional waveforms, our work shows how 
encoding routines can serve as processing units throughout quantum workflows.

The core idea relies on encoding function values (magnitude and phase) into the parameters of controlled rotations on auxiliary qubits. 
Pointwise multiplication of $f(x)$ and $g(x)$ naturally emerges as the coefficient of the $\ket{x}\ket{00}$ state when using two ancillae. 
This forms the basis for implementing convolution via the convolution theorem, by operating in the Fourier domain.

While previous research has explored quantum methodologies that could encompass these operations, our method provides a direct circuit-level construction for these fundamental signal processing operations, which are encompassed within the encoding stage itself.

Integration with the \verb|quantumaudio| package and experiments with audio signals showcase a practical application domain. Audification proved to be a valuable tool for qualitatively assessing the proposed methodology, highlighting the impact of limitations related to encoding
(e.g., limited resolution due to qubit count, normalization effects).
Utilizing massively parallel cluster infrastructure we were further able
to obtain speedups of order linear to the amount of
processing cores used. This integration of the classical simulation
of the generated circuits pushes further 
towards the real-time usage of such tools.

Considering quantum implementations,
the efficiency of the overall process depends on several factors:
\linebreak
1.  The efficiency of the state preparation for the initial functions $f$ (and $g$, or $\widehat{g}$).
2.  The complexity of the controlled rotation gates. Multi-controlled gates can be resource-intensive.
3.  The success probability of post-selecting the desired ancilla state (e.g., $\ket{00}$ for $f(x)g(x)$, which is $\frac{1}{N}\sum_x |f(x)g(x)|^2$).
We note that, for sparse products, or functions with small magnitudes, this can be low, requiring many repetitions or amplitude amplification.

We envision our work to operate within a larger framework,
e.g. as multiple filters, processing audio in a quantum pipeline,
or included in processes that audifiy or sonify quantum phenomena.

\noindent Future work will focus on:\vspace{-0.5\baselineskip}
\begin{itemize}
    \item   Rigorous analysis of the resource requirements
    and success probabilities.
    \item   Investigating the musical and artistic capabilities unlocked by these quantum processing techniques, such as novel sound transformations or synthesis methods.
    \item   Extending experimental validation to the convolution algorithm using audio signals, including designing and testing various quantum filters (kernels $g$).
    \item   Exploring fault-tolerant implementations and noise effects on real quantum hardware.
\end{itemize}

\noindent
Refininng these techniques
will allow us to explore their practical advantages in a quantum computing context.

\bibliographystyle{unsrt}
\bibliography{references}

\end{document}